\documentclass[aps,prl,reprint,amsmath,amssymb,%groupedaddress
superscriptaddress,showpacs,showkeys]{revtex4-1}
\usepackage[none]{hyphenat} %to suppress hyphenation
\usepackage{graphicx}% Include figure files
\usepackage{dcolumn}% Align table columns on decimal point
\usepackage{bm}% bold math
\usepackage{soul} % to cross a text; command \st{....}
\usepackage{color}% print color text
\definecolor{vs}{rgb}{0.1,0.4,0.1}                  % dark green
\newcommand{\del}[1]{}                             % to accept all deletions comment \newcommand{\del}[1]{\textcolor{vs}{\st{#1}}} with % and uncomment this line
\usepackage{verbatim}
\newcommand\wordcount{\verbatiminput{\jobname.sum}}
\usepackage{hyperref}% add hypertext capabilities
\begin{document}

%\preprint{}

\title{Universal tractable model of dynamic resonances and its application to light scattering by small particles}


\author{Michael I. Tribelsky}
\email[Corresponding author:\\]{\mbox{E-mail: }mitribel@gmail.com}
\homepage[Web:~]{https://polly.phys.msu.ru/en/labs/Tribelsky/}
%\thanks{}
\affiliation{%Faculty of Physics,
M. V. Lomonosov Moscow State University, Moscow, 119991, Russia}
\affiliation{National Research Nuclear University MEPhI (Moscow Engineering Physics Institute), Moscow, 115409, Russia}
\affiliation{General Physics Institute RAS, Moscow 119991, Russia}
\affiliation{RITS Yamaguchi University, Yamaguchi, 753-8511, Japan}
\author{Andrey E. Miroshnichenko}
\affiliation{School of Information and Information Technology, University of New South Wales, Canberra, ACT, 2600, Australia}
\email[]{\mbox{E-mail: }andrey.miroshnichenko@unsw.edu.au}
\homepage[Web:~]{http://andreysquare.com/}

\date{\today}

\begin{abstract}
If the duration of the input pulse resonantly interacting with a system is comparable or smaller than the time required for the system to achieve the steady state, transient effects become important. For complex systems, a quantitative description of these effects may be a very difficult problem. We suggest a simple tractable model to describe these phenomena. The model is based on approximation of the actual Fourier spectrum of the system by that composed of the superposition of the spectra of uncoupled harmonic oscillators (normal modes). The physical nature of the underlying system is employed to select the proper approximation.  This reduces the dynamics of the system to tractable dynamics of just a few driven oscillators. The method is simple and may be applied to many types of resonances. As an illustration, the approach is employed to describe the sharp intensive spikes observed in the recent numerical simulation of short light pulses scattered by a cylinder in the proximity of destructive Fano interference [Phys. Rev. A., vol. 100, 053824 (2019)] and exhibits excellent agreement with the numerics.
\end{abstract}

%\pacs{}
%\keywords{}

\maketitle
{\it Introduction}. Various resonant phenomena are of utmost importance in a wide variety of problems ranged from astronomy and cosmology to nuclear and particle physics. Recent advances in laser physics have made possible to generate powerful light pulses whose duration is comparable with the atomic time-scale.

On the other hand, the higher the $Q$-factor of resonance, the longer the transient period for the system to achieve the resonant steady state. During the transient, the manifestation of effects, associated with the resonant excitation of the system may undergo drastic changes. It opens a door to a new discipline --- dynamic resonant phenomena. The discipline has already attracted the attention of explorers, see, for example,\mbox{~\cite{kaldun2016observing,Svyakhovskiy:19,Tribelsky_Mirosh:PRA_2019}.} However, if the system in question is complex enough, the {\it ab initio} description of its transient dynamics is a difficult problem. In this case, a simple yet accurate analytically tractable robust model applicable to a wide range of various resonant systems is badly needed.

For the time being, commonly used models of such a kind are based on the Temporal Coupled-Mode Theory (TCMT)~\cite{louisell1960coupled}. Though the TCMT is a powerful tool indeed, it is not free from disadvantages. First, its applicability conditions imply weak coupling between the modes. Second, the expression of TCMT parameters in terms of the actual parameters of the underlying problem sometimes is not so straightforward. For example, at the application of TCMT to resonant light scattering by a particle, an important parameter is the effective reflection coefficient relating the outgoing wave to the incoming wave in each scattering channel~\cite{Fan:TCM_2010}. However, the implementation of this coefficient requires the balance of the incoming, outgoing and dissipated power. The balance holds at the steady-state scattering solely. Thus, the extension of the TCMT to a non-steady scattering requires essential modifications of the existing theory and remains an open question.

{\it Approach.} For this reason, an alternative approach, free from the mentioned disadvantages of TCMT is highly desirable. Such an approach is introduced in the present Letter. It is quite general and may be applied to a wide variety of linear systems exhibiting oscillatory behavior, whose dynamics are described by second-order in time differential equations. If the systems are continuous so that the corresponding equations are PDEs, they may be reduced to ODEs, e.g., by presenting the solution as a series in spatial eigenfunctions. Then, if a linear system governed by a second-order ODE has an oscillatory dynamics, it is equivalent to coupled harmonic oscillators. The number of oscillators equal to the number of the eigenfrequencies of the system. Next, the linear transformation of the variables to the normal modes makes the oscillators uncoupled. The last step is to fit the steady-state spectrum of the corresponding superposition of the uncoupled normal modes to the actual steady-state spectrum of the original system. This fixes the eigenfrequencies of the oscillators and the amplitude(s) of the drive(s). That is it! As soon as this is done, the superposition of the vibrations of the oscillators presents the temporal evolution of the actual system. In a scene, this approach is similar to the harmonic inversion of time signals and related Prony method in signal processing, where a given signal is fitted by a weighted sum of damped sinusoids, see, e.g, Ref.~\cite{Mandelshtam:JCP:1997,barone1989segmented,roessling2015finite}
However, our method is considerably simpler and often may provide a closed analytical solution describing a complicated dynamic with high accuracy. Moreover, since in contrast to formal mathematical signal processing in our approach the mode selection for the fit is based on the physical properties of the system in question, it reveals the role of different excitations in the system dynamics and shed light on the physical nature of the system as a whole.

This is a sketch of the approach. Let us elaborate it applying to particular cases.

{\it Two coupled oscillators and generalized Fano resonances}. To begin with, we consider the simplest system exhibiting the Fano resonances, namely, two coupled harmonic oscillators~\cite{Joe2006,Tribelsky:RITS}:
\begin{eqnarray}
  & & \ddot{z}_{1} + 2\gamma\dot{z}_{1} +\omega_{01}^2 z_{1} = A(t)\exp[-i\omega t] + \kappa z_2, \label{eq:z1} \\
  & & \ddot{z}_2 + \omega_{02}^2 z_2 = \kappa z_{1}, \label{eq:z2}
  \end{eqnarray}
supplemented with the initial conditions ${z}_{1}(0)=\dot{z}_{1}(0)={z}_{2}(0) =\dot{z}_{2}(0)=0$, where dot stands for $d/dt$. Regarding $A(t)$, for the sake of simplicity here we suppose $A(t)=0$ at $t<0$ and \mbox{$A(t)=A_0$} at \mbox{$t \geq 0$,} where $A_0$ is a constant (generally speaking, complex). According to the procedure described above, we have to focus on the steady-state solution of the problem at $t \rightarrow \infty$. For $z_1$ this solution reads

\begin{equation}\label{eq:z1_steady}
  z_{1s}(t) = %Z_{1s}\exp[-i\omega t],\;\; Z_{1s}=
 - \frac{A_0(\omega^2 - \omega_{02}^2)\exp[-i\omega t]}{(\omega^2-\omega_{02}^2)(\omega^2-\omega_{01}^2 + 2 i \gamma  \omega)-\kappa ^2}.
\end{equation}
It vanishes at $\omega = \omega_{02}$ owing to the mutual cancelation of the drive and the coupling force (an analog of the destructive interference in wave scattering).

Eqs.~\eqref{eq:z1}--\eqref{eq:z2}, as well as its extension to any number of coupled oscillators $N$, are exactly integrable. In terms of the normal modes this solution reads~\cite{LL_mechanics}:
\begin{eqnarray}
  z_{k}(t) = z_{ks}(t)+
 \sum_{n}{\Delta_{kn}C_n e^{-i\hat{\omega}_n t}} \label{eq:solz12},
\end{eqnarray}
where $k=1,2,..N;\; z_{ks}(t)$ are the steady-state solutions for $z_k$ analogous to that given by Eq.~\eqref{eq:z1_steady}; \mbox{$\Delta_{kn}$,} are certain coefficients, and $\hat{\omega}_{n}$ stand for the problem eigenfrequencies, generally speaking complex, given by the poles of the expressions for $z_{ks}$. The arbitrary constants $C_n$ are fixed by the initial conditions.

In the case of Eqs.~\eqref{eq:z1}--\eqref{eq:z2}, $k=1,2$; \mbox{$\Delta_{1n}=1$;} and \mbox{$\Delta_{2n}=\kappa/(\omega_{02}^2-\hat{\omega}_n^2)\equiv (\omega_{01}^2-2i\gamma\hat{\omega}_n-\hat{\omega}_n^2))/\kappa$}. The complex eigenfrequencies of the normal modes have the form \mbox{$\hat{\omega}_n \equiv \pm\omega_{1,2} - i\gamma_{1,2}$}, where $\omega_{1,2}$ and $\gamma_{1,2}$ are supposed to be positive quantities. One should not be confused with the negative signs of the imaginary parts of $\hat{\omega}_n$ --- at the selected time-dependence ($\exp[-i\omega t]$) this sign corresponds to damped oscillations. At small $\kappa$ and $\gamma$  the roots of the characteristic equation, determining the eigenfrequencies, may be readily found by iterations in the vicinity of \mbox{$\omega=\pm\omega_{01}$ and $\omega=\pm\omega_{02}$,} respectively. The first iteration gives rise to:
\begin{eqnarray}
  & &\omega_1 \approx \omega_{01} + \frac{\kappa^2}{2\omega_{01}\left(\omega_{01}^2 - \omega_{02}^2 \right)},\;\;\gamma_1 \approx \gamma; \label{eq:omega_gamma1} \\
  & &\omega_2 \approx \omega_{02} + \frac{\kappa^2\left(\omega_{02}^2-\omega_{01}^2\right)}
  {2\omega_{02}\left[4\gamma^2\omega_{02}^2+\left(\omega_{01}^2-
  \omega_{02}^2\right)^2\right]}, \label{eq:omega2} \\
  & & \gamma_2\approx\frac{\gamma\kappa^2}{4\gamma^2\omega_{02}^2+\left(\omega_{01}^2-
  \omega_{02}^2\right)^2}. \label{eq:gamma2}
\end{eqnarray}
We will not proceed with these trivial calculations. The above expressions have been derived just to make the following important conclusions.

\textit{\textbf{First}}. An expression of the type of Eq. \eqref{eq:solz12}, where the first term in the r.h.s is the steady-state solution, unambiguously determined by the Fourier spectrum of the system and the rest is the sum of the normal modes is the {\it generic\/} solution to a wide class of oscillatory systems.

\textit{\textbf{Second}.} The quantitative description of the transient resonant dynamics of a system may be done {\it without knowledge of the values of the actual parameters of the initial underlying problem}, such as $\gamma, \kappa$, and $\omega_{01,02}$ in Eqs.~\eqref{eq:z1} \eqref{eq:z2}. The only required parameters are the amplitudes and the eigenfrequencies of the normal modes. The latter may be obtained directly from the Fourier spectrum of the steady state of the system.

\textit{\textbf{Third}}. The conventional Fano resonances~\cite{Fano:NC:1935,Fano:PR:1961} correspond to the superposition of two partitions: resonant, whose phase and amplitude sharply depend on the frequency, and background, for which these quantities are constants. In our formalism, it means superposition of vibrations of an oscillator with a finite $Q$-factor with those of an ``oscillator" with the $Q$-factor equals zero.

On the other hand, Eq.~\eqref{eq:solz12} corresponds to the superposition of the vibrations of two oscillators with the {\it finite} $Q$-factor {\it both} ($z_{1s}(t)$ vanishes at $\omega = \omega_{02}$ so that the cancelation of the oscillations is related to the superposition of the two eigenmodes solely). If one of the $\gamma_{1,2}$ is much larger than the other, the mode with the larger $\gamma$ may be treated as the background, while the other should be regarded as resonant. However, when $\gamma_1 \approx \gamma_2$, the division of the modes into the background and resonant becomes meaningless, while the exact cancelation of them at $\omega = \omega_{02}$ remains. We will name this case {\it the generalized Fano resonances}. As we will see below, the mentioned finiteness of the $Q$-factors of both interfering modes is important for the description of the transient dynamics of the Fano resonances.

{\it Dynamics of destructive Fano resonance at light scattering}. To see how the approach may be applied to more complicated cases, we consider its application to the resonant scattering of a laser pulse by a circular infinite dielectric cylinder, whose numerical study is discussed in our recent publication~\cite{Tribelsky_Mirosh:PRA_2019}. In this paper, the direct numerical integration of the complete set of the Maxwell equations is performed for a rectangular pulse with the duration $\tau$ and the carrier frequency $\omega$ (time dependence $\exp[-i\omega t]$) scattered by a lossless cylinder with the base radius $R$ and constant positive refractive index $m$. The normal incidence and TE polarization of the pulse are considered. The well-known exact solution of the problem, see, e.g.,~\cite{Bohren::1998}, is an infinite series of multipoles whose amplitude (so-called, scattering coefficients) read as follows:
\begin{eqnarray}
  a_\ell &=& \frac{mJ_\ell(mq)J^{\prime}_\ell(q) - J_\ell(q)J_\ell'(mq)}{mJ_\ell(mq)H^{(1)\prime}_\ell(q) - H^{(1)}_\ell(q)J_\ell'(mq)},\\ \label{eq:a_ell}
  d_\ell &=& \frac{{2i}/{(\pi q)}}{mJ_\ell(mq)H^{(1)\prime}_\ell(q) - H^{(1)}_\ell(q)J_\ell'(mq)}\label{eq:d_ell}\;,
\end{eqnarray}
where $q = kR$ stands for the size parameter, \mbox{$k=\omega/c$,} $c$ is the speed of light in a vacuum, $J_\ell(z)$ and $H^{(1)}_\ell(z)$ designate the Bessel and Hankel functions, respectively, and prime denotes the derivative over the entire argument of a function. The coefficients $a_\ell$ and $d_\ell$ describe the scattered field outside and within the cylinder, respectively. The coefficients $a_\ell$ and $d_\ell$ are connected by the identity~\cite{Tribelsky_Mirosh_2016}:
\begin{equation}\label{eq:al-dl}
  a_\ell \equiv a_\ell^{(\rm PEC)} - \frac{J^\prime_\ell(mq)}{H^{(1)\prime}_\ell(q)}d_\ell;\; a_\ell^{(\rm PEC)} \equiv \frac{J^\prime_\ell(q)}{H^{(1)\prime}_\ell(q)},
\end{equation}
where and $ a_\ell^{(\rm PEC)}$ is the scattering coefficient of the same cylinder made of the $P$erfect $E$lectric $C$onductor.

Note that Eq.~\eqref{eq:al-dl} has a profound physical meaning. Namely, the basic concept of the conventional Fano resonance in a quantum scattering of a particle~\cite{Fano:NC:1935,Fano:PR:1961} is related to the existence of two possibilities for the particle to be scattered: The first is the collision with the scattering center (elastic or inelastic, such as e.g., ionization), when the particle bypasses the center and goes to infinity (the background scattering). The second is when the particle is trapped by the center so that a quasi-bound state is created. The compound, quasi-bound state %the particle --- the scattering center
lives for a while as a single entity. Then, the particle is released owing to the tunneling through the corresponding potential barrier and again goes to infinity. The full wave function for the particle is a linear superposition of the two partial wave functions describing the two scattering partitions.

Now, let us look at Eq.~\eqref{eq:al-dl} carefully. The term $a_\ell^{(\rm PEC)}$ corresponds to the scattering associated with the surface currents solely when the electromagnetic field does not penetrate the scattering particle, i.e, in this case, the incident wave {\it bypasses} the scatterer. In contrast, the second term in the identity designates the contribution of a resonant localized mode (polariton in dielectrics, plasmon in metals) excited in the particle by the incident wave. In quantum language, the birth of the localized mode is the photon $\rightarrow$ polariton (plasmon) transformation, while the polariton (plasmon) lifetime is determined by the opposite process of its spontaneous transformation into the traveling electromagnetic wave (photon). Now, it is obvious, that the presentation $a_\ell$ in the form of eq.~\eqref{eq:al-dl}, actually, is the division it into the background $\left(a_\ell^{(\rm PEC)}\right)$ and resonant $\left({d_\ell J^\prime_\ell(mq)}/{H^{(1)\prime}_\ell(q)}\right)$ partitions.

To describe the dynamic scattering, we, following Ref.~\cite{Tribelsky_Mirosh:PRA_2019}, introduce the instantaneous scattering cross section $\sigma_{\rm sca}(t)$ as the ratio of the instantaneous total power emitted by the cylinder per unit length of its axis to the intensity of the incident rectangular pulse at $0<t<\tau$. The corresponding scattering efficiency, $Q_{\rm sca}(t)$ is connected with $\sigma_{\rm sca}(t)$ by the usual relation $Q_{\rm sca}(t) = \sigma_{\rm sca}(t)/(2\pi R)$ . The only difference between this and the conventional definition is that we do not perform the averaging of the Poynting vector throughout the field oscillations. Accordingly, the Poynting vector is defined as a real quantity proportional to the vector product of the real parts of the \emph{instantaneous} values of the fields $\mathbf{E(\mbox{$t$},r)}$ and $\mathbf{H(\mbox{$t$},r)}$ in a given point of the space $\mathbf{r}$. Then, for the steady-state scattering~\cite{Tribelsky_Mirosh:PRA_2019}:
\begin{eqnarray}
   \!\!\!\!\!\!\!\!& & Q_{\rm sca}=Q_{\rm sca}^{(0)} + Q_{\rm sca}^{(\rm osc)} = \sum_{\ell=-\infty}^{\infty}\left\{Q^{(0)}_{{\rm sca}\,(\ell)}+ Q_{{\rm sca}\,(\ell)}^{(\rm osc)}\right\},%\; Q_{{\rm sca}\,(\ell)} = Q_{{\rm sca}\,(\ell)}^{(0)} + Q_{{\rm sca}\,(\ell)}^{(\rm osc)},
   \label{eq:Qsca} \\
   \!\!\!\!\!\!\!\!& & Q_{{\rm sca}\,(\ell)}^{(0)}\! =\! \frac{2}{q}|a_\ell(t)|^2;\; Q_{{\rm sca}\,(\ell)}^{(\rm osc)}\! = \! -\frac{i}{q}\!\left[a_\ell^2(t) e^{2ikr}\!-\!c.c.\right]\!\!, \label{eq:Q_through_a}
\end{eqnarray}
where $a_\ell(t) = a_\ell\exp[-i\omega t]$.The same is true for $d_\ell(t)$. Note that \mbox{$a_\ell = a_{-\ell}$,} \mbox{$d_\ell = d_{-\ell}$}~\cite{Bohren::1998}.

Our goal is to recover the dependence $Q_{\rm sca}(t)$ based on its spectrum for the steady-state case. To this end, in accord with our approach, we replace in Eq.~\eqref{eq:al-dl} \mbox{$a_\ell \rightarrow a_\ell(t)$}, \mbox{$d_\ell \rightarrow d_\ell(t)$}. Regarding the $a_\ell(t)$, $d_\ell(t)$, at \mbox{$\ell \neq 0,\pm 2$} they are replaced by the corresponding steady-state quantities, namely  $a_\ell\exp[-i\omega t]$ and $d_\ell\exp[-i\omega t]$, while at \mbox{$\ell = 0,\pm 2$,} they are approximated by the forced vibrations of a harmonic oscillator with the drive $A_0[\theta(t) - \theta(t-\tau)]$. Here \mbox{$A_0=const$} and $\theta(x)$ is the Heaviside step function. The corresponding governing equation is obtained from Eq.~\eqref{eq:z1} at \mbox{$\kappa = 0$}. Its solution, satisfying the initial conditions $z(0)=\dot{z}(0)=0$ is
\begin{eqnarray}
  z(t) &=& A_0 e^{-i\omega t}\times \label{eq:zon}\\
  & & \frac{e^{(i\omega-\gamma)t}\left[
  \left(\gamma-i\omega\right)
  \sin(\omega_{0\gamma}t)+\omega_0
  \cos(\omega_{0\gamma}t)-\omega_{0\gamma}\right]}
  {\left(\omega^2-\omega_0^2+2i\omega\gamma\right)
  \omega_{0\gamma}}, \nonumber
\end{eqnarray}
at $0\leq t \leq \tau$ and
\begin{eqnarray}
  z(t) &=& \frac{e^{-\gamma(t-\tau)}}{\omega_{0\gamma}}
  \Big(\left[\gamma z(\tau)+\dot{z}(\tau)\right]
  \sin\omega_{0\gamma}(t-\tau)+ \nonumber \\
  & & \omega_{0\gamma}z(\tau)\cos\omega_{0\gamma}(t-\tau)\Big), \label{eq:zoff}
\end{eqnarray}
at $t>\tau$. Here $\omega_{0\gamma} \equiv \sqrt{\omega_0^2-\gamma^2}$. Naturally, the values of $A_0,\;\omega_0$ and $\gamma$ are different for different modes and will be defined below during the fitting procedure.

{\it Fitting procedure}. From now on it is convenient to transfer to the dimensionless $\omega$, numerically equal to $q$, i.e., $\omega_{\rm new}=\omega_{\rm old}R/c$ and the corresponding dimensionless time: $t_{\rm new}=t_{\rm old}c/R$. Since in what follows only the dimensionless quantities are in use, the subscript ``new" will be dropped.

In Ref.~\cite{Tribelsky_Mirosh:PRA_2019} the simulation is performed for \mbox{ $m=m_{\rm sim}= 3.125$} and \mbox{ $q=q_{\rm sim}=\omega_{\rm sim}=1.702$.} This choice is done since the pair of $m$ and $q$ corresponds to a local minimum of $Q_{\rm sca}^{(0)}$ associated with the destructive Fano resonances at $\ell=0,\;\pm 2$. At $m=m_{\rm sim}$ they are for \mbox{$\ell = 0$} at \mbox{$q \approx 1.695$} and for \mbox{$\ell = \pm 2$} at \mbox{$q \approx 1.759$.}

The plots of $|a_{0,\pm 2}^{\rm (PEC)}|$ and $|d_{0,\pm 2}|$ are shown in \mbox{Fig.~\ref{fig:F1}}, panels \mbox{(a)--(c).} It is seen that while the $d$-modes have narrow resonant lines, the lines for PEC-modes are broad. Thus, in this case, the Fano resonances are close to the conventional type and the division of the scattered field into the background and resonant partitions is meaningful. Therefore, the transient of the PEC-modes to the steady-state scattering is fast, while the one for the resonant partitions is slow. Accordingly, the approximation of the latter requires maximal accuracy. Bearing this in mind and taking into account that each oscillator has four fitting parameters: Re$[A_0]$, Im$[A_0],\;\omega_0$ and $\gamma$, we adopt the following fitting procedure:
\begin{figure}
  \centering
  \includegraphics[width=\columnwidth]{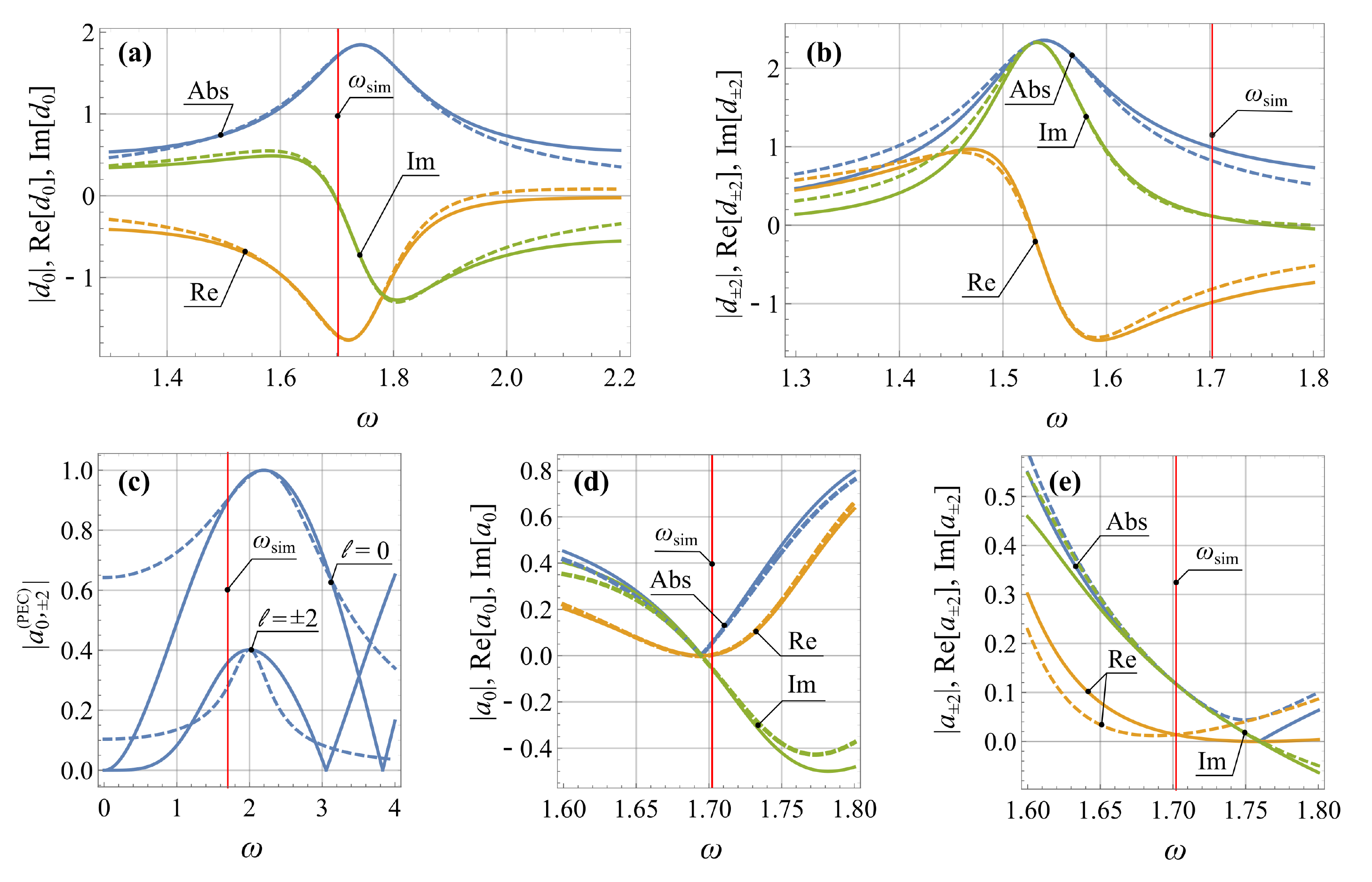}
  \caption{Exact steady-state scattering coefficients (solid lines) and the corresponding fits (dashed lines). }\label{fig:F1}
\end{figure}

\textbf{\textit{d}-modes}. We request that (i) the maximum of modulus of the steady-state profile of the fitting oscillator ($\mathop {\rm Max}\limits_\omega \{ |z_{1s}|\}$, given by Eq.~\eqref{eq:z1_steady} at $\kappa=0$) coincides with the local maximum of $|d_\ell(q)|$ and both are achieved at the same frequency $\omega=q_{\rm max}$; (ii) the phase of the complex quantity $d_\ell(q_{\rm sim})$ is equal to that of the steady-state of the oscillator at the given frequency. The remaining single ``degree of freedom" is employed to fit the linewidth of the oscillator to that of the $|d_\ell(q)|$ profile. The fitting procedure is not unique, however, the final result is rather robust for the specific fitting algorithm. The simplest one, namely just the request of equality of the full widths at half maximum, already gives a good result. More sophisticated fits may improve it, though not very much. Specifically, in our calculations we employed the routine fit of {\it Mathematica} package in the domain \mbox{$1.55 \leq q \leq 1.92$,} whose boundaries are the roots of the equation \mbox{$|d_\ell(q)| = |d_\ell(q_{\rm max})|/2$.} Eventually, this procedure brings about the following values of the fitting parameters:
\begin{eqnarray}
  & & {\rm At}\;\;\ell = 0:\;\;   \gamma \approx 0.101,\; A_0 \approx -0.299 + 0.578i; \label{eq:param_d0} \\
  & & {\rm At}\;\;\ell = \pm 2:\;\;   \gamma \approx 0.064,\; A_0 \approx 0.447 + 0.118i.  \label{eq:param_d2}
\end{eqnarray}
The results are shown in Fig.~\ref{fig:F1}, panels \mbox{(a)--(b)}.

\textbf{\textit{PEC-modes}}. This case is quite different from the previous one because the line shape for the PEC-modes hardly can be accurately approximated with that for a harmonic oscillator, see Fig.~\ref{fig:F1}(c). Fortunately, the effective linewidths of the PEC-modes are high, i.e., their dynamics are fast. It means, that even if the approximation is rough, it gives rise to an error just in the very initial part of the entire transient process, while the PEC-modes arrive at their steady states. After that, the transient is associated with the slow dynamics of the resonant $d$-modes discussed above. We have checked that even the very rough approximation, implying the instant excitation of the PEC-modes into the steady states (corresponding to the conventional Fano resonances with $Q$-factor of the background equals zero) provides rather good agreement with the numerics. Still, it may be considerably improved, if the finiteness of the background $Q$-factor is taken into account. To this end, we approximate the PEC-modes profiles with those for harmonic oscillators too. Once again, for each approximation, we request the equality of the maxima of the modula of the fitting and actual profiles achieved at the same maximizing frequency. Two more remaining conditions we employ to provide the equality at \mbox{$q=q_{\rm sim}$} of the values of the actual complex $a_\ell$ following from the exact solution, Eq.~\eqref{eq:a_ell} to the r.h.s. of Eq.~\eqref{eq:al-dl}, where $a_\ell^{\rm (PEC)}$ and $d_\ell$ are replaced by the corresponding fitting harmonic oscillator profiles. This procedure fixes all four fitting parameters for each of the profiles in question. The corresponding numerical values of the parameters are as follows:
\begin{eqnarray}
  & & {\rm At}\;\;\ell = 0:\;\; \gamma \approx 0.857,\; A_0 \approx 3.906 - 1.039i; \label{eq:param_d0} \\
  & & {\rm At}\;\;\ell = \pm 2:\;\; \gamma \approx 0.265,\; A_0 \approx -0.108 - 0.416i,  \label{eq:param_d2}
\end{eqnarray}
see Fig.~\ref{fig:F1}, panels (c)--(e).

\begin{figure}
  \centering
  \includegraphics[width=\columnwidth]{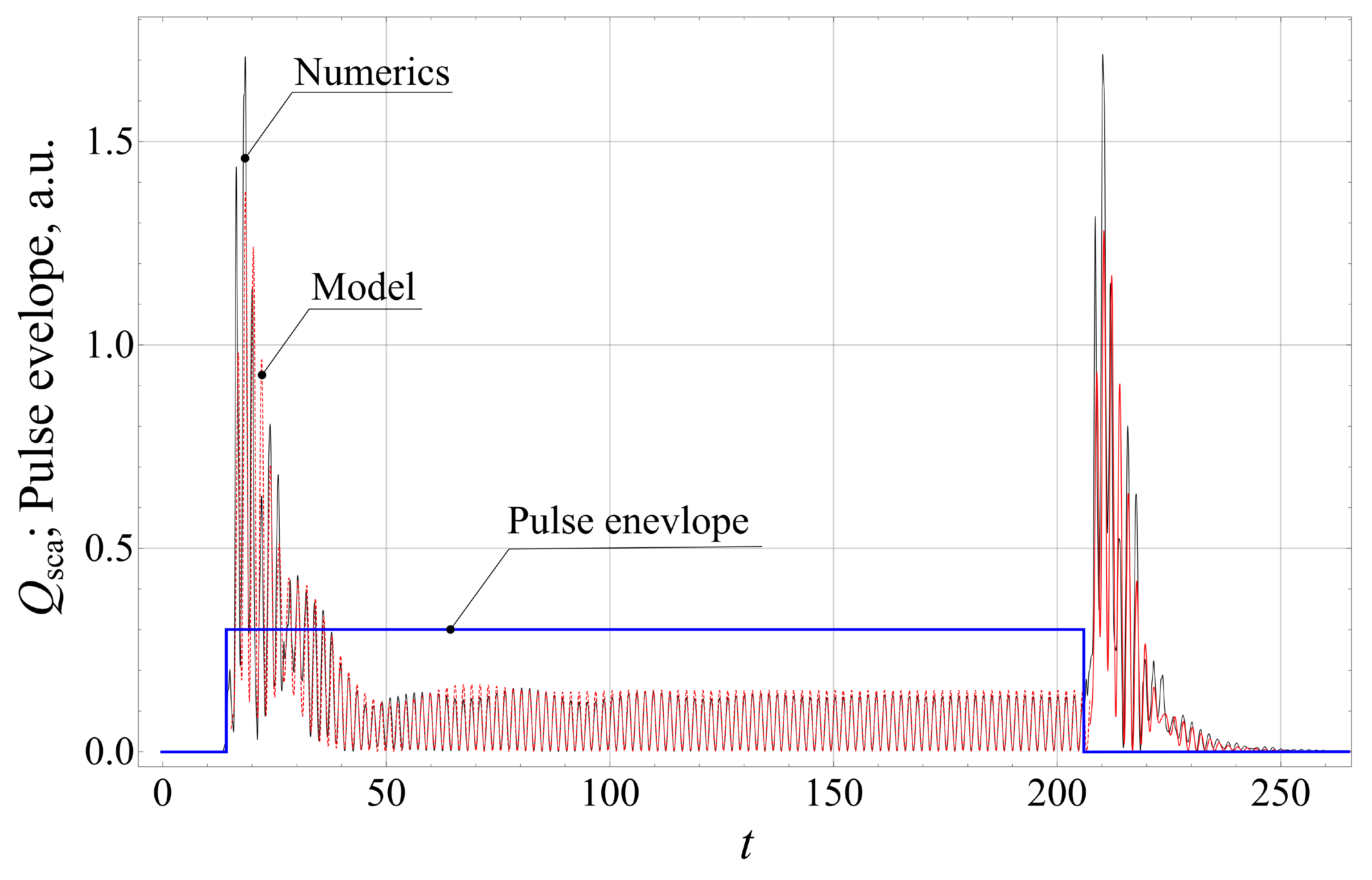}
  \caption{Comparison of $Q_{\rm sca}(t)$ obtained by the direct numerical integration of the complete set of the Maxwell equations~\cite{Tribelsky_Mirosh:PRA_2019} and in our model; $t$ is dimensionless, normalized on $R/c$; $\tau = 192$. The beginning of the scattering is delayed for the time required for light to pass from the inlet to the cylinder and from it to the monitors. The envelope of the incident pulse is shifted accordingly so that the front edge of it coincides with the one for the scattered radiation.}\label{fig:F2}
\end{figure}

Comparison of the developed model with the obtained values of its parameters and the results of the direct numerical integration of the complete set of the Maxwell equations presented in Ref.~\cite{Tribelsky_Mirosh:PRA_2019} is shown in Fig.~\ref{fig:F2}. A qualitative explanation of the sharp spikes in the scattered radiation situated behind the leading and trailing edges of the incident pulse is given already in Ref.~\cite{Tribelsky_Mirosh:PRA_2019}: They are caused by the violation of the balance between the resonant and background partitions during the transient due to the difference in the characteristic times of their dynamics. The disbalance destroys the mutual cancelation of the partitions. The uncompensated partition (background at the leading edge and resonant at trailing) is observed as the spike.

Our approach makes it possible to describe this effect qualitatively. In our view, bearing in mind the simplicity of the model and the complexity of the initial underlying problem, the agreement between them is quite impressive.

{\it Conclusions}. Summarizing the obtained results, we may say that the developed simple approach indeed is a powerful tool to describe the dynamic resonant phenomena in discrete and continuous systems of different nature. Despite its simplicity, the approach exhibits high accuracy even if the problem it is applied to is rather complex. The great advantages of the approach are in the following: (i) it may be applied to systems with an arbitrary number of coupled modes and does not have any restriction on the coupling strength; (ii) all parameters of the corresponding models are readily obtained from the Fourier spectrum of the steady state of the initial underlining system, even when the spectrum is obtained experimentally and does not have any analytical description.

{\it Acknowledgements}. The authors are very grateful to Boris Y. Rubinstein for his valuable help in symbolic computer calculations. M.I.T. acknowledges the financial support of the Russian Foundation for Basic Research (Project No. 20-02-00086) for
the analytical study, the Moscow Engineering Physics Institute Academic Excellence Project (agreement with the Ministry of Education and Science of the Russian Federation of 27 August 2013, Project No. 02.a03.21.0005) for the modeling of the resonant light scattering and computer simulation, and the contribution of the Russian Science Foundation (Project No. 19-72-30012) for the provision of user facilities. The work of A.E.M. was supported by the Australian Research Council
and the University of New South Wales Scientia Fellowship.

%%%%%%%%%%%%%%%%%%%%%%%%%%%%%%%%%%%%%%%%%%%%%%%%%%%%%%%%%%%%%%%%%%%%%%%%%%%%%%%%%%
\bibliography{Dynamic_resonanses} %Put the filename of your bib file here (without extension .bib

\end{document}